\documentclass[
 reprint,
 superscriptaddress,
 amsmath,
 amssymb,
 aps,
 prm,
 floatfix,
 nofootinbib,
]{revtex4-2}

\usepackage{graphicx}
\usepackage{bm}
\usepackage[dvipsnames]{xcolor}
\usepackage{siunitx}
\usepackage[version=4]{mhchem}
\usepackage{xspace}
\usepackage[
    colorlinks=true,
    linkcolor=NavyBlue,  
    citecolor=NavyBlue,  
    urlcolor=gray,       
]{hyperref}

\newcommand*{\fref}[1]{Fig.~\ref{#1}}
\newcommand*{\tref}[1]{Table~\ref{#1}}

\newcommand*{\sref}[1]{Section~\ref{#1}}

\newcommand{\sifull}[0]{Supplemental Material\xspace}
\newcommand{\sishort}[0]{SM\xspace}

\begin{document}

\preprint{}
\title{Guided Synthesis of EMT Zeolites by Machine Learning}

\author{Emmanuel A.\ Olanrewaju}
\author{Santosh Adhikari}
\author{Zhiyin Niu}
\author{Michael Nikolaou}
\author{Jeremy C.\ Palmer}
\author{Jeffrey D.\ Rimer}
\author{Mingjian Wen}
\email{mjwen@uestc.edu.cn}
\thanks{Present address: Institute of Fundamental and Frontier Sciences, University of Electronic Science and Technology of China, Chengdu, 611731, China}
\affiliation{William A. Brookshire Department of Chemical and Biomolecular Engineering, University of Houston, Houston, TX, 77204, USA}

\begin{abstract}
    Zeolites are microporous crystalline materials with diverse frameworks, widely used in industrial applications such as petroleum refining and molecular separation.
    Unlike most zeolites, EMT can be synthesized under mild conditions (at low temperatures and without the use of organic structure-directing agents), making it attractive for cost-effective and environmentally sustainable production.
    However, the specific synthesis conditions that selectively produce EMT rather than similar frameworks like FAU are not yet well established.
    In this work, we develop machine learning (ML) models to guide the discovery of synthesis conditions for EMT zeolites.
    Our dataset comprises 174 experimental synthesis attempts, recording reaction time,
    temperature, silica and alumina sources,  Si/Al stoichiometric ratio, and other synthesis parameters.
    We apply both classical ML methods and pretrained foundation models to predict zeolite framework outcomes from these synthesis parameters.
    Feature importance analysis identifies critical parameters for EMT formation, validating known synthesis principles.
    Leveraging the ML models, we explore the synthesis space and identify six promising new conditions for EMT formation.
    Experimental validation confirms EMT crystallization in five cases, including two with
    Si/Al stoichiometric ratios outside the training dataset's range.
    Evaluation on independent literature-reported synthesis conditions further demonstrates the generalizability of the model.
    This work demonstrates a data-driven approach to accelerating zeolite synthesis, closing the loop between ML prediction and experimental validation.
\end{abstract}

\maketitle

\begingroup
    \renewcommand{\thefootnote}{}
    \footnotetext{
    \textcopyright{} 2026 American Physical Society.
    This is the accepted manuscript of the following article:
    E.~A.\ Olanrewaju \textit{et al.},
    ``Guided synthesis of EMT zeolites by machine learning,''
    \textit{Phys. Rev. Materials} \textbf{10}, 083801 (2026).
    The final published version is available at
    \href{https://doi.org/10.1103/v2yp-ylxm}{https://doi.org/10.1103/v2yp-ylxm}.
    This manuscript version is posted with permission for non-commercial
    scholarly use.}
\endgroup

\section{Introduction}

Zeolites are crystalline materials containing pores and channels arranged in highly ordered three-dimensional frameworks.
Their unique pore sizes and channel geometries, which can be chemically tuned, make them invaluable for industrial processes like petroleum refining (where they help convert crude oil into gasoline) and molecular separation (where they selectively filter molecules based on size and shape)\cite{davis2002ordered,primo2014zeolites,mintova2015nanosized,li2017applications}.
The synthesis of a specific zeolite framework depends on a complex interplay of parameters
such as atomic composition (e.g., \ce{SiO2} and \ce{Al2O3} molar amounts and stoichiometric ratio),  reaction conditions (e.g., time and temperature), and even precursor sources (e.g.,
fumed vs colloidal silica sources)~\cite{li2023regulation,ma2022machine}.
No unique synthesis route exists for a given zeolite framework; variations in synthesis parameters can yield the same crystalline phase, demonstrating the flexibility of the synthesis process \cite{fletcher2015intrinsic,ghojavand2023flexibility,ohsuna2004fine,yang2010machine,dib2022role}.
For example, EMT zeolites, typically synthesized at low temperatures of 30--60~\textdegree C with Si/Al ratios around 2 \cite{ng2012capturing,maldonado2013controlling}, have been successfully dealuminated (post-processed) at elevated temperatures of 450--600~\textdegree C to obtain Si/Al ratios ranging from 4.5 to 52 \cite{morin1997dealumination}.
Despite this flexibility, small changes in synthesis parameters can result in completely different zeolite frameworks \cite{moliner2013towards,oleksiak2014synthesis,muraoka2019linking}.
For example, at low temperature, minor variations in temperature or reaction time can lead to a transition from EMT to alternative frameworks such as FAU or SOD \cite{maldonado2013controlling}.
This dual behavior reveals fundamental gaps in understanding the factors governing zeolite synthesis.

Computational methods have been extensively employed to study zeolites.
Molecular simulations using classical force fields have successfully modeled molecular adsorption and transport processes within zeolites  \cite{slater2001atomistic,psofogiannakis2015reaxff,tesson2018classical,misturini2022molecular}.
First-principles calculations based on density functional theory (DFT) have provided accurate predictions of electronic structure, framework thermodynamic stability, and zeolite-catalyzed reactions \cite{filippone1995structural,astala2004density,shi2014ab,rey2017ab,kim2021accelerated,ducamp2022prediction}.
Modeling the zeolite synthesis process, however, still remains infeasible.
Classical force fields require extensive system-specific parameterization \cite{gaillac2020speeding}, and are further limited by accessible time and length scales, typically capturing only microsecond-scale processes in complex, multi-component synthesis environments of synthesis gel.
DFT, while more accurate, incurs prohibitive computational costs for the relevant length and time scales involved in zeolite crystallization~\cite{ma2022machine,wu2025ai}.

More recently, machine learning (ML) techniques have been applied to investigate zeolite synthesis.
Extensive databases of zeolites have been constructed using data mining and natural language processing to systematically extract synthesis information from the literature~\cite{yan2009database,jensen2019machine,pan2024zeosyn}.
Using such databases, ML models have been developed for mapping the synthesis--structure--property relationship \cite{evans2017predicting, Moliner2019, Conroy2022, SchwalbeKoda2023,carr2009machine,yang2009identifying,raman2023forecasting,li2023machine,peng2025molecular}.
These studies have provided insights into the factors governing zeolite synthesis, such as the effects of synthesis parameters on product yield~\cite{Conroy2022} and  inorganic structure-directing agents~\cite{SchwalbeKoda2023}.
Despite the progress and success, most existing work in this domain employs ML models to categorize synthesis conditions based on the resulting zeolite frameworks.
However, the reliability of these predictions often remains unverified, primarily for two reasons.
First, the adopted ML methods inherently contain errors and uncertainties, particularly when trained on limited datasets~\cite{cohen2008insights, Wen2022, dai2024uncertainty}.
Second, and more critically, the proposed synthesis conditions are rarely experimentally validated~\cite{ma2022machine, gandhi2022machine}.

In this work, we integrate data analytics, ML methods, and experiments to study the synthesis conditions that govern EMT formation.
Using an in-house dataset of 174 experimental synthesis conditions targeting EMT zeolites, we develop ML models to predict crystallization outcomes.
Comprehensive evaluation of multiple algorithms, including both classical statistical ML methods~\cite{hastie2009elements} and the recently introduced TabPFN foundation model \cite{hollmann2025accurate},
reveals comparable performance across all methods.
The ML models successfully identify key synthesis parameters such as temperature, Si/OH stoichiometric ratio, Si/Al stoichiometric ratio, and Si molar amount that play a significant role in the formation of EMT zeolites.

We also leverage the developed ML models to explore synthesis conditions that favor EMT formation.
This is accomplished by first generating new synthesis conditions in a dimension-reduced latent space, followed by screening these conditions based on both the predicted likelihood of forming EMT frameworks (as determined by the ML models) and additional criteria, such as the uniqueness of each condition.
Subsequent synthesis experiments demonstrate that five of the top six predicted conditions result in a pure EMT phase.
Thus, the ML models are found to have a success rate of 83\%.
This value is significantly larger than the baseline prevalence in the training data set, where only 32\% of the conditions yielded pure EMT. Notably, while all synthesis mixture Si/Al ratios in the training data exceeded 2.5, two of the five successful conditions exhibited lower ratios of 1.77 and 2.3.
This demonstrates that our ML approach can identify synthesis conditions that extrapolate beyond the parameter space of the training data, uncovering previously unexplored regions of the chemical space.

Finally, we benchmark the model against literature-reported synthesis conditions to assess its external generalizability.
This benchmark shows that the model reliably predicts EMT outcomes for literature cases that remain close as well as distant from the training domain, including cases with NaOH far outside the original training range.

\section{Methods}
\label{sec:datasets:methods}

\begin{table*}[tbh!]
    \centering
    \caption{Experimental synthesis parameters for EMT zeolites.
        X is one of: fumed silica, Ludox-AS40, Ludox-SM30, \ce{Na2SiO3} and tetraethyl orthosilicate (TEOS);
        Y is either \ce{Al(OH)3} or \ce{NaAlO2}.
        Two types of features are used: numerical (N), which can have values in a continuous range,
        and categorical (C), which takes discrete values.
    }
    \label{tab:features}
    \begin{tabular}{llc}
        \hline
        Feature     & Explanation                                                                           & Type \\
        \hline
        Time        & Synthesis time (hour)                                                                 & N    \\
        Temperature & Synthesis temperature (\textdegree C)                                                 & N    \\
        Si          & Molar amount of \ce{SiO2}                                                             & N    \\
        Al          & Molar amount of Al cations                                                            & N    \\
        NaOH        & Molar amount of NaOH                                                                  & N    \\
        \ce{H2O}    & Molar amount of \ce{H2O}                                                              & N    \\
        Si/OH       & Stoichiometric ratio between \ce{SiO2} and NaOH                                       & N    \\
        Si/Al       & Stoichiometric ratio between \ce{SiO2} and Al cation                                  & N    \\
        Si size     & \ce{SiO2} particle size (nm) in the silica source                                     & N    \\
        Si:X        & Reagent (X) that provides \ce{SiO2} to the synthesis mixture                          & C    \\
        Al:Y        & Reagent (Y) that provides \ce{Al} cation to the synthesis mixture                     & C    \\
        Pd:Si       & Whether the \ce{SiO2} reagent is pre-dissolved before adding to the synthesis mixture & C    \\
        Pd:Al       & Whether the Al reagent is pre-dissolved before adding to the synthesis mixture
                    & C                                                                                            \\
        \hline
    \end{tabular}
\end{table*}

\subsection{Dataset}
\label{sec:data:cleaning}

The in-house dataset is built around canonical OSDA-free (organic structure-directing agent free) EMT synthesis routes reported in the literature~\cite{ng2012capturing,ng2015emt,maldonado2013controlling}.
We retained the key features that define these routes (alkaline, OSDA-free synthesis from sodium-based silica and aluminum precursors) and then deliberately broadened the synthesis space around them.
Compared to the literature routes, our dataset spans a much wider range of compositions, crystallization temperatures and times, and precursor choices, so that the ML model can learn how EMT formation depends on synthesis parameters beyond the narrow window of the literature conditions.
A side-by-side comparison of the literature routes and the in-house parameter ranges is provided in Table~S1 of the \sifull (\sishort)~\cite{supplemental}.

The raw experimental data was cleaned by checking for duplicates, estimating missing values, and removing information that can only be determined post-synthesis, such as crystal morphology and pH.
See \sishort for detailed data cleaning procedures.
After cleaning, the final dataset consisted of 174 samples with 13 synthesis parameters (i.e., features) and a label indicating the final synthesis outcome.
The features include composition information (e.g., Si and Al molar amounts),
precursor type (e.g., fumed silica or Ludox-AS40 as Si source), synthesis time, and temperature, among others.
\tref{tab:features} lists all features, along with their data type.
The synthesis outcome label consists of three distinct values:
`pure-EMT' for zeolites with a pure EMT phase;
`non-EMT' for outcomes in amorphous or solution forms where no EMT phase was observed;
and `hybrid-EMT' for zeolites exhibiting intergrowth of EMT with other phases such as FAU.
Out of the 174 samples, there are 57 pure-EMT, 69 hybrid-EMT, and 48 non-EMT.
A pair plot showing the distributions of the features is provided in Fig.~S1 in the \sishort{}.

\subsection{Model Training}
\label{sec:model:training}

Given the relatively small dataset, we employed classical statistical ML algorithms (Random Forest \cite{breiman2001random}, XGBoost \cite{chen2016xgboost}, and CatBoost \cite{prokhorenkova2018catboost}) to develop classification models to map the features to the labels.
We also used TabPFN~\cite{hollmann2025accurate}, a recently proposed foundation model pre-trained on massive synthetic data to address classification problems with small data, which is well-suited to our case.

The features were preprocessed before being fed to the ML algorithms.
Numerical features with continuous values (e.g., synthesis time) were normalized to have a mean of zero and a standard deviation of one to ensure that the scale of different features does not bias the learning process.
Categorical features that take discrete values (e.g., the reagent that provides \ce{SiO2}) were encoded using a one-hot scheme. 

For all models trained in this work,
the dataset was randomly split into three subsets for training, validation, and testing, with a ratio of 8:1:1.
The model parameters were optimized using the training set, hyperparameters were determined based on model performance on the validation set, and performance metrics are reported on the test set in the main text.
Results on the training and validation sets are provided in the \sishort.
To reduce the dependency of the model on a particular data split, we trained 35 models, each with a different data split.
The reported performance metrics represent the average across all models, with error bars indicating the standard deviation of these measurements.

\section{Results and Discussion}
\label{sec:results}

\subsection{Importance of Synthesis Parameters}
\label{sec:feat:importance}

\begin{table}[tbh!]
    \caption{Performance of various models on the ternary classification.
        The error is obtained as the standard deviation of predictions from multiple models, each trained on a different data split.
    }
    \label{tab:test:metrics:ternary}
    \begin{tabular}{lcccc}
        \hline
                  & XGBoost         & Random Forest   & CatBoost        & TabPFN          \\
        \hline
        Accuracy  & $0.76 \pm 0.07$ & $0.75 \pm 0.09$ & $0.76 \pm 0.08$ & $0.74 \pm 0.09$ \\
        Precision & $0.79 \pm 0.07$ & $0.78 \pm 0.09$ & $0.79 \pm 0.07$ & $0.76 \pm 0.09$ \\
        Recall    & $0.76 \pm 0.07$ & $0.75 \pm 0.09$ & $0.76 \pm 0.08$ & $0.74 \pm 0.09$ \\
        $F_1$     & $0.76 \pm 0.07$ & $0.75 \pm 0.09$ & $0.76 \pm 0.08$ & $0.74 \pm 0.09$ \\
        \hline
    \end{tabular}
\end{table}

We first benchmark XGBoost~\cite{chen2016xgboost}, Random Forest~\cite{breiman2001random}, CatBoost~\cite{prokhorenkova2018catboost}, and TabPFN~\cite{hollmann2025accurate} by developing ternary classifiers that predict synthesis outcomes (pure‑EMT, hybrid‑EMT, or non‑EMT) from synthesis parameters.
Details of model training are provided above in \sref{sec:model:training}, and their performance metrics, including accuracy, precision, recall, and $F_1$ score, on the test set are summarized in \tref{tab:test:metrics:ternary}.
All four ML models achieve an $F_1$ score of about 0.75, and no model performs significantly better than the others.
This is also true for accuracy, precision, and recall.
XGBoost is selected for subsequent analysis.

\begin{figure}[h!]
    \centering
    \includegraphics[width=1\columnwidth]{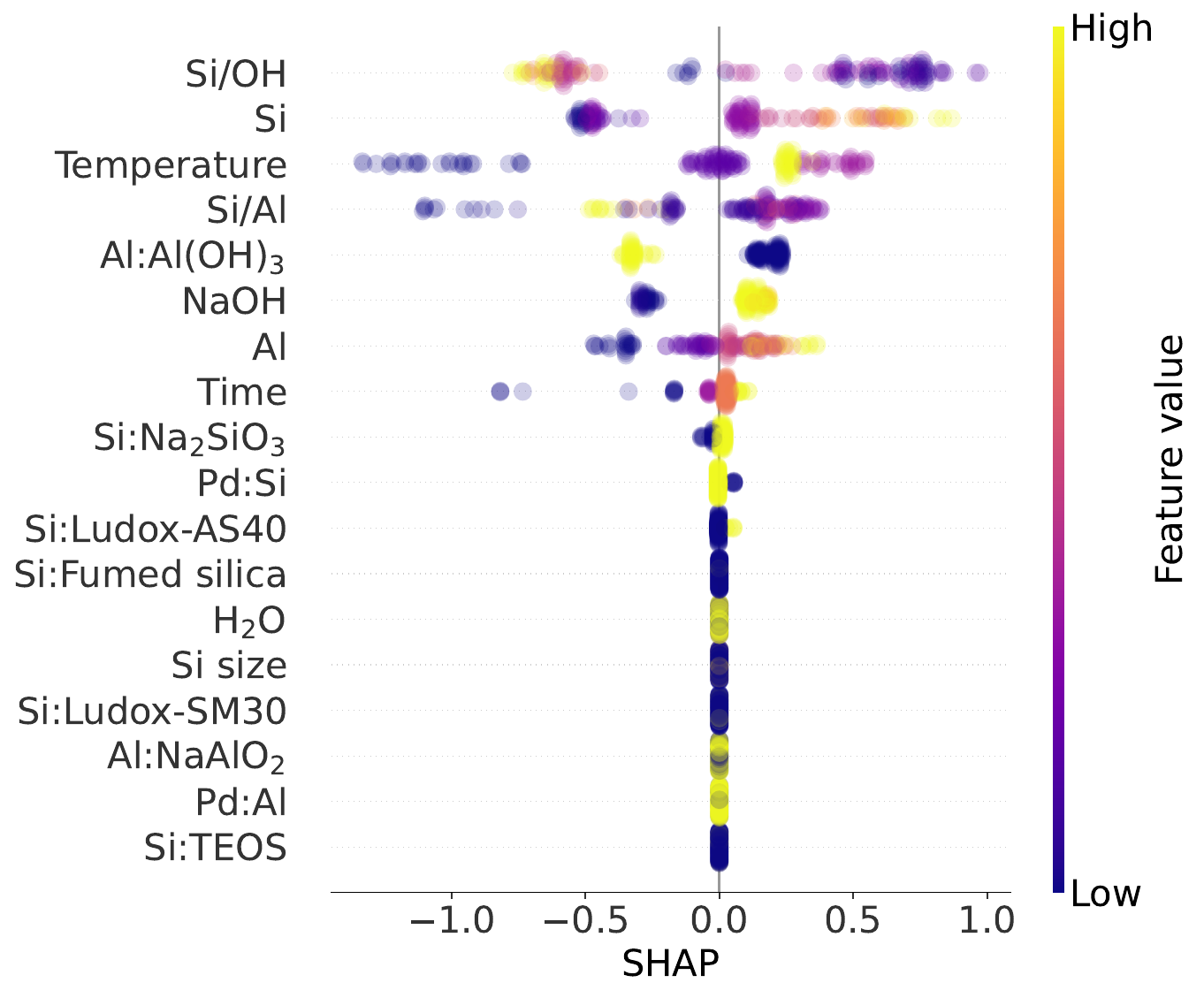}
    \caption{Feature importance analysis using SHAP for the ternary classification model.
        The horizontal spread of SHAP values indicates feature importance: a wider distribution signifies greater influence.
        The 18 features displayed (compared to 13 in \tref{tab:features}) result from one-hot encoding, which expands categorical features into multiple dimensions.
    }
    \label{fig:feat:imp:multiclass}
\end{figure}

Using the XGBoost model, we perform feature importance analysis to  identify key synthesis parameters that control EMT formation.
Feature importance analysis was conducted using both the mean decrease in impurity (MDI) method~\cite{hastie2009elements} and the Shapley Additive Explanations (SHAP) method~\cite{lundberg2017unified}.
MDI quantifies feature importance by measuring the total reduction in impurity (e.g, entropy) brought by each feature across all trees in the ensemble~\cite{hastie2009elements}, while SHAP employs concepts from cooperative game theory to attribute the contribution of each feature to the model's prediction~\cite{lundberg2017unified}.
Moreover, the sign of a SHAP value (positive or negative) indicates whether a feature pushed the model's specific prediction above or below the average baseline prediction, which can provide deeper insight into the model's decision-making process.
Both SHAP analysis (\fref{fig:feat:imp:multiclass}) and MDI (Fig.~S2, \sishort{}) identify the Si/OH ratio, Si/Al ratio, temperature, and molar amounts of Si, NaOH, and Al as the most influential features.
This agreement underscores the importance of these parameters in EMT synthesis, despite differences in their relative rankings between the two methods.

\begin{figure}[bth!]
    \centering
    \includegraphics[width=0.95\columnwidth]{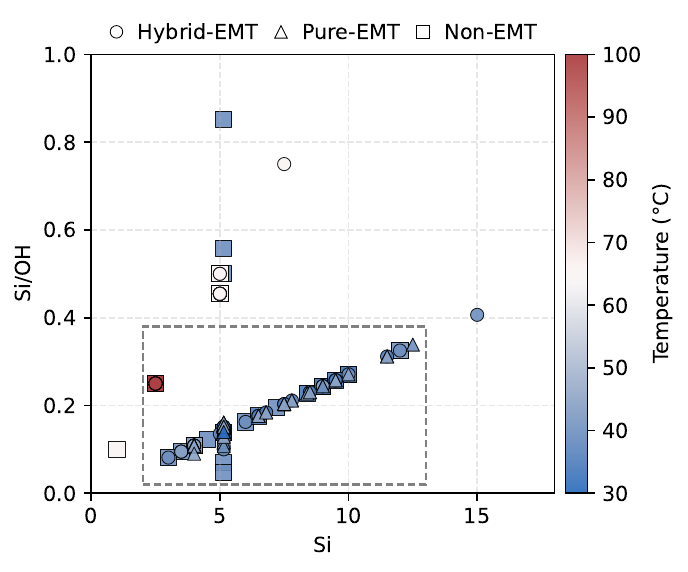}
    \caption{Distribution of synthesis outcome labels with respect to the three most important features.
        These features are \ce{SiO2} molar amount, Si/OH stoichiometric ratio, and temperature.
        All pure-EMT samples lie inside the dashed box.
    }
    \label{fig:dataset:SiOH_Si}
\end{figure}

Although feature importance analysis confirms that synthesis outcomes are governed by multiple parameters, distinct regions in the parameter space correspond to specific zeolite types.
This relationship is illustrated in \fref{fig:dataset:SiOH_Si}, which shows the outcome labels plotted against the three most influential features: Si molar amount, Si/OH ratio, and temperature.
Although pure-EMT samples coexist with hybrid-EMT and non-EMT, they are found exclusively within a specific regime of Si molar amounts and Si/OH ratios (dashed box in \fref{fig:dataset:SiOH_Si}).
The absence of pure-EMT samples outside this region demonstrates that its crystallization is confined to certain synthesis conditions.

\subsection{A Simplified and Focused Model}
\label{sec:binary:model}

\begin{figure}[tbh!]
    \centering
    \includegraphics[width=0.8\columnwidth]{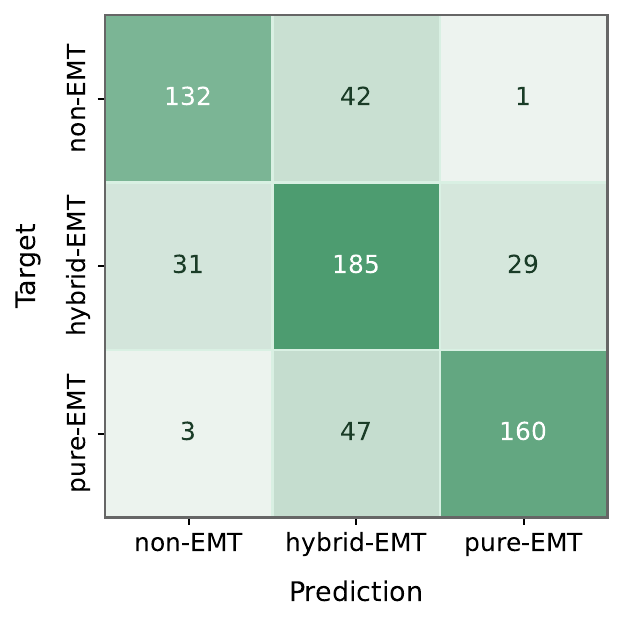}
    \caption{Confusion matrix for the XGBoost ternary classification model.
    }
    \label{fig:confusion:matrix:multiclass}
\end{figure}

A closer examination of the ternary classification model in \sref{sec:feat:importance} reveals that prediction errors are primarily associated with the hybrid-EMT class.
As shown in \fref{fig:confusion:matrix:multiclass}, only three pure-EMT samples are misclassified as non-EMT, and only one non-EMT sample is misclassified as pure-EMT.
This stands in stark contrast to the hybrid-EMT class, which exhibits substantial misclassifications with both pure-EMT and non-EMT.
We posit that because the hybrid-EMT class represents mixed-phase products, its synthesis conditions have a larger overlap with those of both pure-EMT and non-EMT, leading to a higher rate of misclassification.
Since our primary objective is to identify conditions that yield pure-EMT zeolites, we can simplify the problem by merging the hybrid-EMT and non-EMT classes into a single class, called `Others'.
This results in a binary classification of `pure-EMT' versus `Others'.

\begin{figure*}[tbh!]
    \centering
    \includegraphics[width=0.9\textwidth]{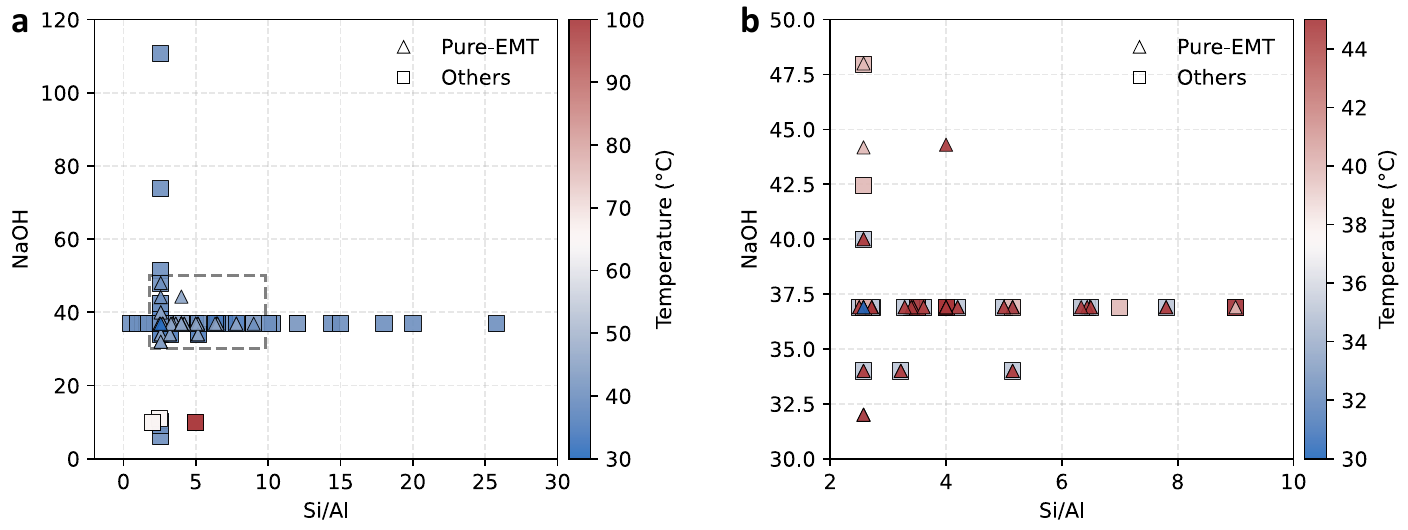}
    \caption{Distribution of binary synthesis outcome labels with respect to Si/Al stoichiometric ratio, NaOH molar amount, and temperature.
        (a) Distribution of all 174 samples in the dataset.
        (b) Zoomed-in view of the dashed region in panel a, which contains all pure-EMT samples.
    }
    \label{fig:dataset:distribution}
\end{figure*}

\begin{table}[tbh!]
    \centering
    \caption{Binary classification results using the reduced dataset of 101 samples.
        The reported values are averaged over predictions from multiple models, each trained on a different data split.
        Therefore, the mean $F_1$ score is not generally equal to the harmonic mean of the average precision and average recall, nor is it guaranteed to lie between them.
        The error is obtained as the standard deviation of  predictions from multiple models.
    }
    \label{tab:test:metrics:binary_101}
    \begin{tabular}{lcccc}
        \hline
                  & XGBoost         & Random Forest   & CatBoost        & TabPFN          \\
        \hline
        Accuracy  & $0.81 \pm 0.12$ & $0.81 \pm 0.13$ & $0.79 \pm 0.13$ & $0.80 \pm 0.13$ \\
        Precision & $0.81 \pm 0.13$ & $0.81 \pm 0.15$ & $0.79 \pm 0.14$ & $0.80 \pm 0.14$ \\
        Recall    & $0.79 \pm 0.13$ & $0.80 \pm 0.15$ & $0.79 \pm 0.14$ & $0.79 \pm 0.14$ \\
        $F_1$     & $0.79 \pm 0.13$ & $0.79 \pm 0.15$ & $0.78 \pm 0.14$ & $0.79 \pm 0.14$ \\
        \hline
    \end{tabular}
\end{table}

\begin{figure}[h!]
    \centering
    \includegraphics[width=1.0\columnwidth]{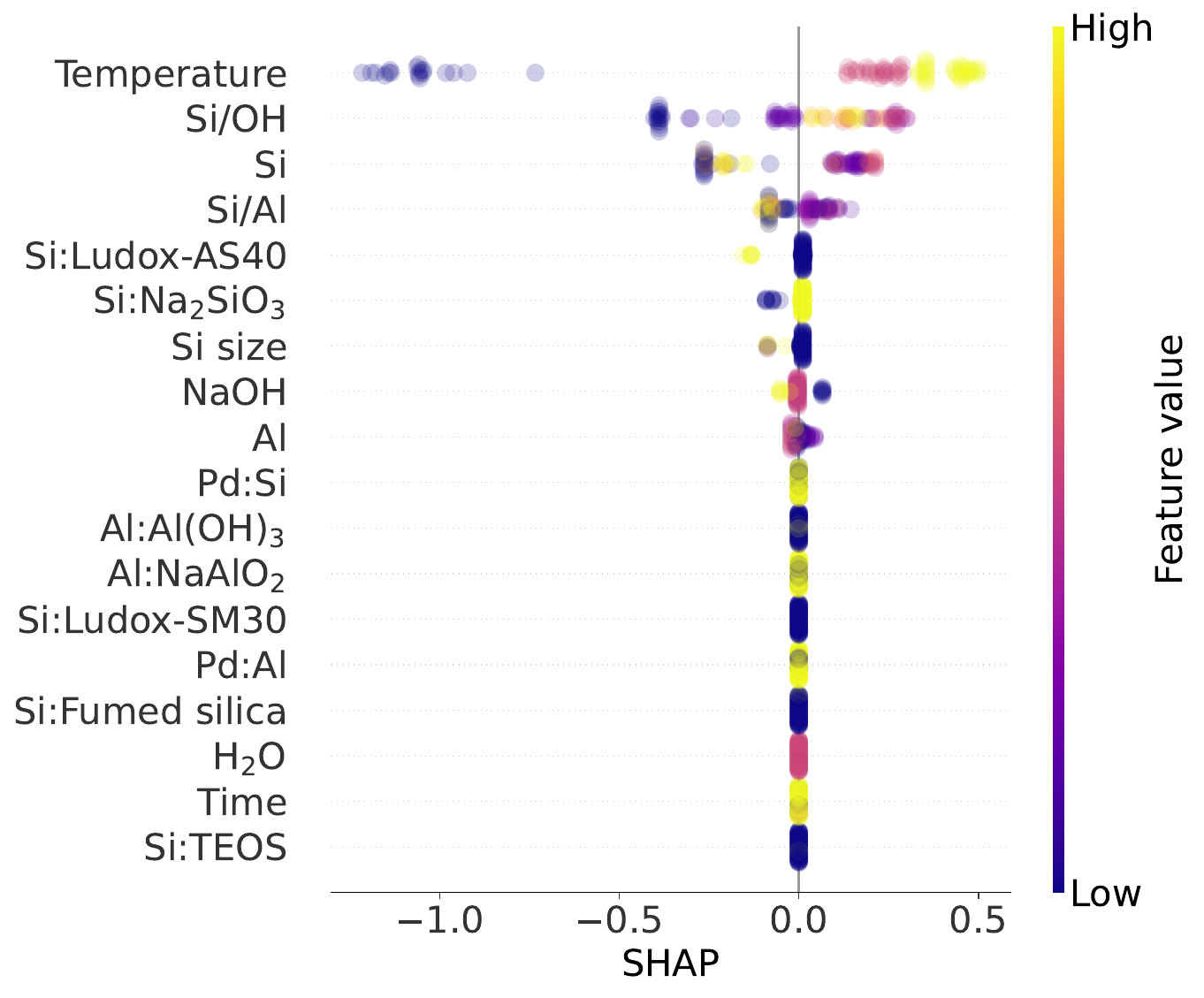}
    \caption{Feature importance analysis using SHAP for the binary classification model.
        The 18 features displayed (compared to 13 in \tref{tab:features}) result from one-hot encoding, which expands categorical features into multiple dimensions.
    }
    \label{fig:shap:binary}
\end{figure}

The task can be further simplified by focusing on the region of the parameter space most relevant to EMT formation.
As discussed above (see \fref{fig:dataset:SiOH_Si}), pure-EMT crystallizes only within specific regions in the parameter space.
For the binary classification task, we identified that the Si/Al ratio and NaOH molar amount are effective at delineating the pure-EMT region (\fref{fig:dataset:distribution}).
All 57 pure-EMT samples are located within the bounds of 2 $\leq$ Si/Al $\leq$ 10 and 30 $\leq$ NaOH $\leq$ 50.
This region contains a total of 101 samples with 57 pure-EMT and 44 Others.
Using these two features for filtering is more effective than using the top two features from the importance analysis (Si molar amount and Si/OH ratio in \fref{fig:dataset:SiOH_Si}), which define a larger region containing 147 samples.
Owing to feature correlations, constraining Si/Al and NaOH also implicitly restricts other parameters.
For example, the temperature within this focused region is confined to 30--45~\textdegree C, a significant reduction from the full dataset's range of 30--100~\textdegree C.

Using the filtered dataset of 101 samples, we developed binary classification models to distinguish pure-EMT from Others.
Similar to the ternary classification, the XGBoost, Random Forest, CatBoost, and TabPFN models demonstrate comparable performance on the test set across all classification metrics, as summarized in \tref{tab:test:metrics:binary_101} (results on the training and validation sets are provided in the \sishort).
At first glance, the performance of the binary classification models may appear only marginally better than that of the ternary classification models presented in \tref{tab:test:metrics:ternary}.
However, it is important to note that in this binary classification task, 73 of the 174 samples in the original dataset were excluded because they can be trivially classified as `Others'.
This is because their synthesis conditions lie outside the defined ranges for the Si/Al ratio and NaOH molar amount.
Taking this into account, the effective performance of the binary classification model is significantly improved.
For example, it achieves an accuracy of 89\%, compared to 76\% for the ternary classification.

SHAP feature importance analysis was also performed for the binary classification model, as shown in \fref{fig:shap:binary}.
Compared to the ternary classification (\fref{fig:feat:imp:multiclass}), the top four most influential features remain the same: temperature, Si/OH and Si/Al stoichiometric ratios, and Si molar amount.
This further confirms their critical role in determining EMT zeolite formation.
For a binary classification model, the SHAP values can be directly interpreted as the contribution of each feature to predicting the classification outcome~\cite{lundberg2017unified}.
In this context, positive SHAP values indicate that a feature value favors pure-EMT formation, while negative SHAP values suggest the opposite.
\fref{fig:shap:binary} shows that, in general, positive SHAP values are associated with high temperature and high Si/OH ratio, suggesting that larger values of the two features can promote pure-EMT formation.
For Si molar amount and Si/Al ratio, high SHAP values are associated with both high and low feature values, indicating a more complex relationship with EMT formation.

\subsection{Discovery of New Synthesis Conditions}
\label{sec:new:synthesis}

\begin{figure*}[tbh!]
    \centering
    \includegraphics[width=0.9\textwidth]{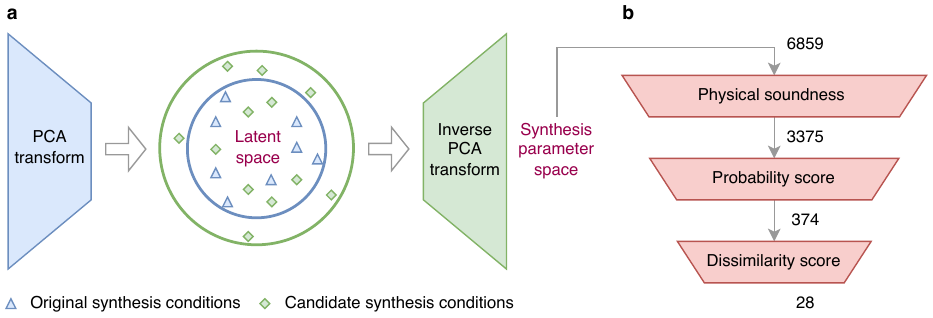}
    \caption{A two-stage pipeline for discovering new synthesis conditions for EMT zeolites.
        (a) New candidate synthesis conditions are proposed in a dimension-reduced latent space obtained via principal component analysis (PCA).
        The blue circle indicates the boundary of the parameter space covered by the dataset, whereas the green circle represents the expanded parameter space explored by the proposer.
        (b) The proposed candidate synthesis conditions are screened to identify the most promising ones.
    }
    \label{fig:procedure}
\end{figure*}

\begin{table*}[tbh!]
    \centering
    \caption{Experimentally verified synthesis conditions proposed by the ML pipeline.
        ``Probability'' is the prediction confidence score from the ML model, and ``Dissimilarity'' is the Euclidean distance of the synthesis condition to the dataset.
        ``Outcome'' indicates the observed zeolite framework after synthesis, but not the label predicted by the ML model.
        Seven features are listed explicitly.
        The remaining six features are the same for all six samples:
        Pd:Si (yes), Pd:Al (yes), Si size (2), Si:\ce{Na2SiO3}, Al:\ce{NaAlO2}, \ce{H2O} (450).
    }
    \label{tab:new:synthesis}
    \begin{tabular}{cccccccccc}
        \hline
        NaOH  & Al   & Si   & Si/OH & Si/Al & Time (hours) & Temperature ($^{\circ}$C) & Probability & Dissimilarity & Outcome    \\
        \hline
        37.39 & 2.08 & 5.27 & 0.14  & 2.32  & 72.11        & 41.39                     & 0.90        & 4.98          & pure EMT   \\
        36.76 & 1.12 & 5.14 & 0.14  & 4.90  & 71.85        & 41.43                     & 0.88        & 4.65          & pure EMT   \\
        36.65 & 1.65 & 5.40 & 0.15  & 3.61  & 72.58        & 41.56                     & 0.88        & 0.67          & pure EMT   \\
        36.93 & 1.26 & 4.15 & 0.11  & 3.69  & 72.32        & 42.21                     & 0.87        & 3.90          & pure EMT   \\
        36.59 & 2.93 & 7.58 & 0.21  & 1.77  & 73.18        & 40.45                     & 0.86        & 1.79          & pure EMT   \\
        36.99 & 1.04 & 5.35 & 0.15  & 5.29  & 71.29        & 41.06                     & 0.93        & 4.65          & hybrid EMT \\
        \hline
    \end{tabular}
\end{table*}

\begin{figure}[tbh!]
    \centering
    \includegraphics[width=0.95\columnwidth]{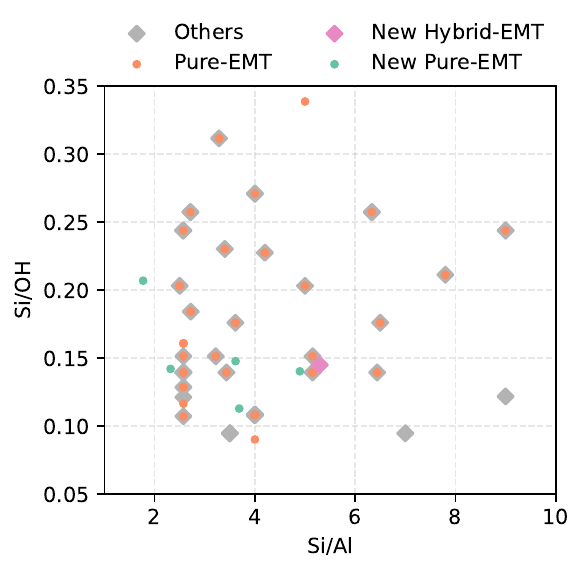}
    \caption{Distribution of the six experimentally verified synthesis conditions proposed by the ML model.
        They correspond to the new pure-EMT and new hybrid-EMT (purple diamond and green circles).
        Samples from the training data are also included (orange circles and gray diamonds).
        The number of training data appears to be less than 101 since many of them overlap in this two-dimensional plot.
    }
    \label{fig:Pair:plot:2feat}
\end{figure}

With the trained binary ML classifier,  we aim to discover new synthesis conditions for synthesizing pure EMT zeolites.
This is achieved through a two-stage pipeline consisting of a ``synthesis parameters proposer'' and a ``synthesis feasibility evaluator'', as illustrated in Figure~\ref{fig:procedure}.
First, the proposer generates new synthesis conditions that could potentially yield pure-EMT zeolites.
Next, the evaluator examines the feasibility of the conditions using the trained ML classifier, supplemented by additional validation checks.
Finally, the most promising candidate synthesis conditions are selected for experimental synthesis and validation.

The synthesis parameters proposer aims to generate new synthesis conditions both within and beyond the parameter bounds of the original dataset.
This can be accomplished through methods such as grid search or random sampling of features beyond their observed ranges.
However, this approach faces two challenges.
A grid search across all 13 features produces an exponentially large number of candidates;
subsequent evaluation of these candidates becomes computationally intensive (see Section~S5 in \sishort{} for a time estimate).
More critically, the features exhibit strong correlations (see Fig.~S1 in \sishort{}), and thus most candidates from grid search or random sampling may be chemically unrealistic or thermodynamically infeasible.

We address these challenges by proposing candidate synthesis conditions in a dimension-reduced latent space obtained via principal component analysis (PCA).
This was carried out as follows (\fref{fig:procedure}a).
First, we transformed the 13-dimensional feature vectors of all data into a three-dimensional latent space using PCA by retaining the first three principal components.
Next, for each of the three principal components we sampled 19 evenly spaced values spanning from 30\% below the observed minimum to 30\% above the observed maximum.
Altogether, 6859 new candidate synthesis conditions were created.
Finally, each candidate in the latent space was mapped back to the original 13-dimensional synthesis parameter space via the inverse PCA transform.

Three filtering criteria were then applied to the PCA-generated populate to identify the most promising candidate synthesis conditions: physical soundness, likelihood of producing pure EMT zeolites, and distinctiveness from known conditions in the dataset (Figure~\ref{fig:procedure}b).
Some candidate conditions contained unphysical parameters, such as negative Si/Al ratios.
After removing these invalid cases, the remaining 3375 candidate conditions were evaluated using the XGBoost classifier developed in Section~\ref{sec:binary:model} to predict their likelihood of yielding pure EMT zeolites.
Rather than using the binary yes/no classification, we examined the continuous probability score predicted by the model, which provides more granular information.
A higher probability score indicates a greater likelihood of forming pure EMT zeolite.
We retained candidates with probability scores greater than 0.85, which reduced the total to 374 candidates.
We then computed Euclidean distances of each of the 374 candidates to all samples in the original dataset and used the minimum distance as the ``dissimilarity score'' $\tilde d$.
This score quantifies the novelty of each candidate relative to known conditions;
a higher value indicates a more novel synthesis condition. Of the 374 conditions, 346 were found to have dissimilarity score below 0.5 and were discarded, yielding a final pool of 28 promising candidates (see Section~S6 in \sishort{} for post-processing details).

From the 28 candidate synthesis conditions, we selected the six candidates with the highest predicted probabilities for experimental synthesis (see Section~S7 in \sishort{} for experimental details).
The predicted synthesis parameters for these six candidates are listed in \tref{tab:new:synthesis}.
Remarkably, five of the six candidates successfully produced pure-EMT zeolites, corresponding to an 83\% success rate.
The remaining candidate yielded a hybrid-EMT phase.
This success rate represents a 2.6-fold improvement over the 32\% success rate achieved by the experience-guided methods used to generate the original dataset.
The structural identity of the resulting EMT samples was confirmed by powder X-ray diffraction against the IZA reference EMT pattern, and scanning electron microscopy of a representative sample showed submicron aggregates of EMT nanocrystallites, consistent with what is typically observed for EMT prepared under OSDA-free conditions~\cite{ng2012capturing,ng2015emt,maldonado2013controlling} (Section~S8 in \sishort{}).

Upon closer examination, we found that two of the five pure-EMT samples were synthesized under extrapolated conditions.
Although all proposed synthesis parameters differ from those in the dataset, most fall within the dataset's observed ranges.
For example, the Si/OH ratios for the six experimentally verified conditions range from 0.11 to 0.21, compared to 0.09 to 0.34 in the original dataset (\fref{fig:Pair:plot:2feat}).
Nevertheless, two conditions have Si/Al ratios of 1.77 and 2.3, which fall below the dataset's lower bound of 2.5.
This observation indicates that our ML-based pipeline can successfully extrapolate beyond the training data to discover novel synthesis conditions.
Since Si/Al ratio is a key determinant of synthesis outcome (Figures~\ref{fig:feat:imp:multiclass} and \ref{fig:shap:binary}), correctly predicting out-of-range Si/Al ratios represents a nontrivial and noteworthy achievement.

\subsection{External Validation on Literature-Reported Syntheses}
\label{sec:new:validation}

The experimental validation in \sref{sec:new:synthesis} confirms the model on new in-house syntheses, including two recipes beyond the training Si/Al range.
A more stringent test of generalizability is whether the model transfers to syntheses performed by other groups, where synthesis conditions such as precursor choices and reaction protocols can differ from those in our in-house dataset.
To probe this, we compiled 16 EMT zeolite synthesis conditions from the literature~\cite{wendelbo1999tumbling,wu1995synthesis,dougnier1995emt,georgieva2015control,ng2012capturing,ng2015emt}.
The full dataset, literature compilation strategy, the procedure for imputing descriptors not reported in the original papers, and the model predictions are described in Section~S9 of the \sishort{}.
Because some descriptors had to be imputed, the literature validation should be read as a partially imputed external evaluation across reported EMT-forming conditions rather than a fully independent test across all model inputs.

\begin{figure}[bth!]
    \centering
    \includegraphics[width=0.95\columnwidth]{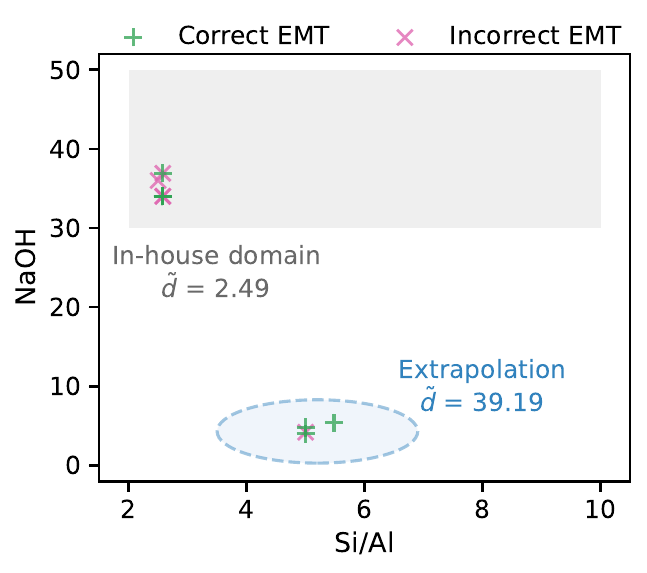}
    \caption{Model predictions for literature-reported EMT synthesis conditions in the NaOH--Si/Al phase space.
        The gray region marks the NaOH and Si/Al envelope of the in-house dataset; the blue region marks extrapolation beyond that envelope.
        The median dissimilarity score $\tilde{d}$ of the literature-reported EMT zeolites in each region is annotated.
        $\tilde d$ measures the distance from a synthesis condition to the in-house dataset (see \sref{sec:new:synthesis} for its definition).
    }
    \label{fig:validation}
\end{figure}

It is shown in \fref{fig:dataset:distribution} that pure-EMT samples in the in-house dataset fall within a narrow range of NaOH and Si/Al values.
Here, we use these two descriptors to organize the literature EMT samples, which fall both within and beyond the NaOH and Si/Al ranges of the in-house dataset (\fref{fig:validation}).
The gray region marks the in-house NaOH--Si/Al envelope, where the literature EMT samples have a median dissimilarity score of $\tilde{d}=2.49$, which lies within the $\tilde{d}\approx 0.7$--$5$ range of the in-house cases we validated experimentally.
In this region, the model correctly predicts 8 of 12 EMT cases (67\%).
The blue region marks literature EMT samples that fall outside the in-house NaOH--Si/Al envelope, with NaOH values below 10.
This is considered an extrapolation regime, as the median dissimilarity score of these cases is $\tilde{d}=39.19$, which is approximately 14 times larger than that of the literature cases in the in-house region.
The model correctly predicts 3 of 4 EMT cases (75\%) in this region.

The higher success rate in the extrapolation region than in the in-house region may seem counterintuitive.
A closer look at the four low-NaOH cases shows that the differences between the correctly and incorrectly classified ones do not appear to align with NaOH or Si/Al, but rather with temperature and crystallization time: the three correct cases use $90$--$115\,^\circ$C and $168$--$624$ h, whereas the misclassified Dougnier case~\cite{dougnier1995emt} uses $72\,^\circ$C and only $24$ h (Table~S7, \sishort{}).
This is broadly consistent with the feature importance analysis of the binary classifier (\fref{fig:shap:binary}), in which temperature is the most important feature and higher temperature tends to favor EMT formation.
That said, the small sample size (4 cases) makes the 75\% rate sensitive to individual predictions and should not be over-interpreted.
Even so, the model correctly identifies the majority of literature-reported EMT cases in both regions, with success rates above the 32\% EMT incidence observed during the manual generation of the in-house dataset.

Overall, the model's performance across both regions suggests that it does not simply memorize the in-house recipes but captures broader relationships among synthesis descriptors, such as the temperature dependence noted above.
A more definitive assessment, however, will require evaluation on a larger set of literature cases.

\section{Conclusion}

We developed a closed-loop framework integrating ML and experimental synthesis to accelerate the discovery of conditions for EMT zeolite formation.
By collecting and systematically analyzing 174 OSDA-free synthesis attempts, we identified temperature, Si/OH ratio, Si/Al ratio, and Si amount as the most critical parameters governing EMT crystallization.
Our binary ML classification model, trained on a refined subset of the data, successfully guided the experimental synthesis of five (out of six) EMT zeolites under new conditions, achieving a much higher success rate than the experience-guided approach.
Notably, two successful conditions extrapolate beyond the knowledge contained in the dataset.
Evaluation against external literature further confirms that the model captures transferable relationships among synthesis descriptors beyond the in-house recipes, correctly predicting the majority of reported EMT cases both within and beyond the training stoichiometric range.

This study demonstrates that data-driven approaches, when coupled with rigorous experimental validation, can significantly improve the effectiveness of zeolite synthesis optimization.
The approach of proposing synthesis conditions in a dimension-reduced space followed by a multi-criteria screening provides a generalizable workflow that can be applied to other zeolite frameworks and, more broadly, to other crystalline materials whose synthesis depends on a complex interplay of compositional and reaction variables.
Looking forward, an iterative active learning strategy could systematically improve model accuracy with minimal experimental effort.
This process would involve incorporating newly validated outcomes and using them to retrain models and guide subsequent experiments.
We anticipate that this iterative approach will enable efficient navigation of the vast synthesis parameter space and accelerate the discovery of new zeolite frameworks.
While this work focuses on classifying synthesis outcomes, future efforts could explore the use of regression models to predict quantities such as crystal size and yield, thus enabling multi-objective optimization toward economical and environmentally sustainable synthesis routes.
Beyond synthesis itself, the long-term phase stability of metastable frameworks like EMT, including their resistance to phase transformation (e.g., to FAU) and amorphization under aging or thermal treatment, is a critical practical consideration that warrants dedicated investigation in future work.

\section*{Data and Code Availability}

The raw and cleaned experimental synthesis data, along with the code for model training and data analysis are available at the GitHub repository: \url{https://github.com/Olanrewajuemmanuelabiodun/ml_EMT_synthesis}.

\section*{Author contributions}
E.A.O. and S.A.: model development, data analysis, writing - original draft, and writing - review.
Z.N. and J.D.R.: data curation and writing - review.
M.N. and  J.C.P.: writing - review.
M.W.: project conceptualization, data analysis, writing - original draft, writing - review, and supervision.

\section*{Conflicts of interest}
There are no conflicts of interest to declare.

\section*{Acknowledgements}

JDR and JCP gratefully acknowledge funding from Department of Energy,
Office of Basic Energy Sciences, Materials Science Division
(Award DE-SC0021384) and additional support from the Welch Foundation (Grants E-1882, E-1794, and Award V-E-0001).
Computational resources were provided by the Research Computing Data Core at the University of Houston and the Hefei Advanced Computing Center.

%

\end{document}


\title{Supplemental Material:\\ Guided Synthesis of EMT Zeolites by Machine Learning}

\author{Emmanuel A.\ Olanrewaju}
\author{Santosh Adhikari}
\author{Zhiyin Niu}
\author{Michael Nikolaou}
\author{Jeremy C.\ Palmer}
\author{Jeffrey D.\ Rimer}
\author{Mingjian Wen}
\email{mjwen@uestc.edu.cn}
\thanks{Present address: Institute of Fundamental and Frontier Sciences, University of Electronic Science and Technology of China, Chengdu, 611731, China}
\affiliation{William A. Brookshire Department of Chemical and Biomolecular Engineering, University of Houston, Houston, TX, 77204, USA}

\maketitle

\clearpage

\section{Data cleaning}
\label{sec:si:data:cleaning}

The raw experimental data, as initially collected, were pre-processed to ensure consistency and completeness for analysis.
The raw data consists of 20 different experimental measurements from 199 attempts to synthesize EMT zeolites.

\begin{figure}[tbh!]
    \centering
    \includegraphics[width=1.10\textwidth]{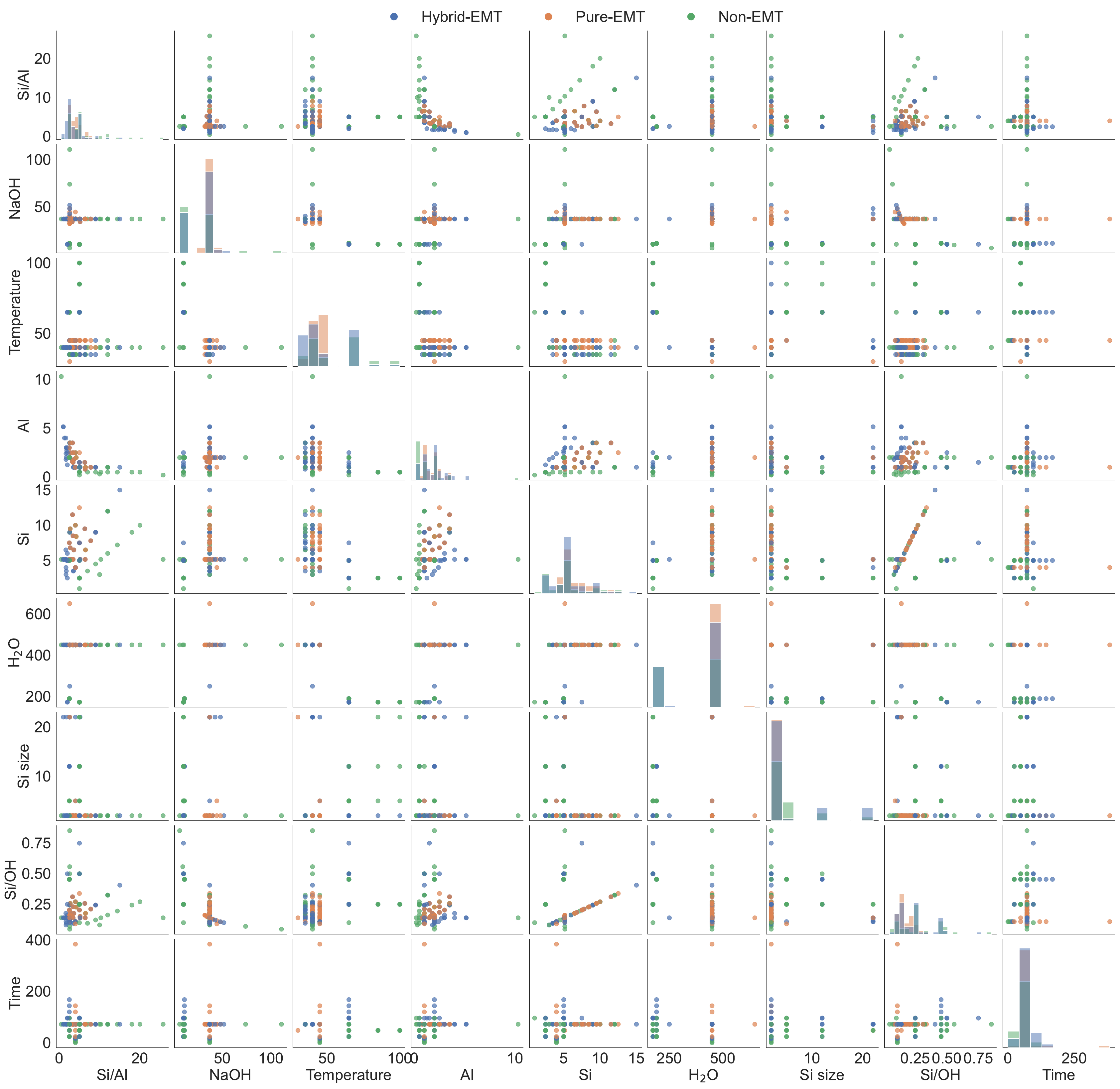}
    \caption{Pair plot of the distributions of dataset features.
        Continuous features are included, whereas categorical ones are not.
    }
    \label{fig:pair:plot}
\end{figure}

The data cleaning involves several steps.
\begin{itemize}
    \item Repeated entries were removed to avoid duplication.
    \item Missing values were imputed using the mean of the respective feature.
    \item Post-synthesis features, which are unavailable before an experiment is performed, were removed. This includes features like pH value and $V_{\text{micore}}\,(\mathrm{cm}^3\!/\mathrm{g})$ (micropore volume). This ensures that the model can be used for predictive purposes.
\end{itemize}
The cleaned dataset consists of 174 data points with 13 features and a label indicating the synthesized outcome: \verb|pure-EMT|, \verb|non-EMT|, or \verb|hybrid-EMT|.
\fref{fig:pair:plot} shows a pair plot of the distribution of the features.

\begin{table}[h!]
    \centering
    \caption{Comparison between canonical OSDA-free EMT synthesis routes reported in the literature~\cite{ng2012capturing,ng2015emt,maldonado2013controlling} and the in-house dataset of 174 samples used in this work.
        Molar amounts of NaOH, Al, Si, and \ce{H2O} are reported relative to a unit cell of the gel composition.
        ``Literature routes'' values correspond to representative conditions reported in the references; ``In-house dataset'' values are the full ranges observed across the 174 samples.}
    \label{tab:dataset:comparison}
    \begin{tabular}{lll}
        \hline
        Synthesis parameter                         & Literature routes \cite{ng2012capturing,ng2015emt,maldonado2013controlling} & In-house dataset                                     \\
        \hline
        NaOH                                        & 36--37                                                                      & 6--111                                               \\
        Al                                          & $\sim$2                                                                     & 0.2--10.3                                            \\
        Si                                          & 5--5.15                                                                     & 1--15                                                \\
        \ce{H2O}                                    & 217--240                                                                    & 173--650                                             \\
        Si/Al                                       & 2.5--2.6                                                                    & 0.5--25.8                                            \\
        Crystallization time (h)                    & 28--36                                                                      & 0--384                                               \\
        Crystallization temperature (\textdegree C) & 28--30                                                                      & 30--100                                              \\
        Si precursor particle size (nm)             & not specified                                                               & 2--22                                                \\
        Si source                                   & \ce{Na2SiO3}                                                                & \shortstack[l]{\ce{Na2SiO3}, Ludox-AS40, Ludox-SM30, \\ fumed silica, TEOS} \\
        Al source                                   & \ce{NaAlO2}                                                                 & \ce{NaAlO2}, \ce{Al(OH)3}                            \\
        OSDA                                        & None                                                                        & None                                                 \\
        \hline
    \end{tabular}
\end{table}

\section{Feature vector}

A distinct characteristic of our dataset is the presence of numerous categorical features, such as the silicon (Si) and Aluminium (Al) source used in zeolite synthesis.
In Table~I of the main text, categorical features are indicated by ``C'' in the ``Type'' column.
Among the 174 zeolites in our dataset, each reports one of five Si sources: fumed silica, Ludox-AS40, Ludox-SM30, sodium silicate (\ce{Na2SiO3}), or tetraethyl orthosilicate (TEOS).
To encode this categorical information, we applied one-hot encoding, representing the Si source using five binary columns, each corresponding to one of the five possible sources.
If a zeolite was synthesized using fumed silica, the first column was assigned a value of 1, while the remaining columns were set to 0.
Similarly, if the Si source was Ludox-SM30, the third column was assigned 1, with all others set to 0, and so forth.
Boolean features such as Pre-dissolution of Si and Pre-dissolution of Al, which contained either ``yes'' or ``no'' responses, were encoded such that ``yes'' was assigned a value of 1 and ``no'' was assigned 0.
With these transformations, the input to our model consists of a fixed-length feature vector of size 18, representing the 13 features listed in Table~I of the main text.

\section{Comparison of classification models}

To identify the most suitable model for our classification task, we compared the performance of Random Forest (RF), XGBoost, CatBoost, and TabPFN.
We randomly partitioned the dataset into training (80\%), validation (10\%), and test (10\%) subsets. The models were independently trained and evaluated on these partitions.
\tref{tab:train:metrics:ternary}, \tref{tab:val:metrics:ternary}, \tref{tab:train:metrics:binary_101}, and \tref{tab:val:metrics:binary_101} summarize the classification metrics (accuracy, precision, recall, and $F_1$-score).
The results show that XGBoost achieves competitive performance compared to the other models. Given its strong performance and widespread use in scientific applications, we selected XGBoost for all subsequent analyses in this work.
\begin{table}[h!]
    \centering
    \caption{Classification performance on the train set for all models for ternary classification.}
    \begin{tabular}{lcccc}
        \toprule
                  & \textbf{XGBoost} & \textbf{RF}     & \textbf{CatBoost} & \textbf{TabPFN} \\
        \midrule
        Accuracy  & $0.95 \pm 0.01$  & $0.87 \pm 0.02$ & $0.91 \pm 0.02$   & $0.91 \pm 0.02$ \\
        Precision & $0.95 \pm 0.01$  & $0.88 \pm 0.02$ & $0.92 \pm 0.01$   & $0.92 \pm 0.02$ \\
        Recall    & $0.95 \pm 0.01$  & $0.87 \pm 0.02$ & $0.91 \pm 0.02$   & $0.91 \pm 0.02$ \\
        $F_1$     & $0.95 \pm 0.01$  & $0.87 \pm 0.02$ & $0.91 \pm 0.02$   & $0.92 \pm 0.02$ \\
        \bottomrule
    \end{tabular}
    \label{tab:train:metrics:ternary}
\end{table}

\begin{table}[h!]
    \centering
    \caption{Classification performance on the validation set for all models for ternary classification.}
    \begin{tabular}{lcccc}
        \toprule
                  & \textbf{XGBoost} & \textbf{RF}     & \textbf{CatBoost} & \textbf{TabPFN} \\
        \midrule
        Accuracy  & $0.77 \pm 0.12$  & $0.76 \pm 0.12$ & $0.78 \pm 0.11$   & $0.75 \pm 0.10$ \\
        Precision & $0.80 \pm 0.11$  & $0.78 \pm 0.11$ & $0.81 \pm 0.10$   & $0.78 \pm 0.10$ \\
        Recall    & $0.77 \pm 0.13$  & $0.76 \pm 0.12$ & $0.78 \pm 0.12$   & $0.75 \pm 0.10$ \\
        $F_1$     & $0.77 \pm 0.13$  & $0.76 \pm 0.12$ & $0.78 \pm 0.12$   & $0.75 \pm 0.10$ \\
        \bottomrule
    \end{tabular}
    \label{tab:val:metrics:ternary}
\end{table}

\begin{table}[h!]
    \centering
    \caption{Classification performance on the train set for all models for binary classification for 101 datasets.}
    \begin{tabular}{lcccc}
        \toprule
                  & \textbf{XGBoost} & \textbf{RF}     & \textbf{CatBoost} & \textbf{TabPFN} \\
        \midrule
        Accuracy  & $0.84 \pm 0.03$  & $0.87 \pm 0.02$ & $0.91 \pm 0.01$   & $0.92 \pm 0.03$ \\
        Precision & $0.85 \pm 0.03$  & $0.87 \pm 0.02$ & $0.92 \pm 0.01$   & $0.92 \pm 0.03$ \\
        Recall    & $0.83 \pm 0.03$  & $0.86 \pm 0.02$ & $0.91 \pm 0.02$   & $0.92 \pm 0.03$ \\
        $F_1$     & $0.84 \pm 0.03$  & $0.86 \pm 0.02$ & $0.91 \pm 0.02$   & $0.92 \pm 0.03$ \\
        \bottomrule
    \end{tabular}
    \label{tab:train:metrics:binary_101}
\end{table}

\begin{table}[h!]
    \centering
    \caption{Classification performance on the validation set for all models for binary classification for 101 datasets.}
    \begin{tabular}{lcccc}
        \toprule
                  & \textbf{XGBoost} & \textbf{RF}     & \textbf{CatBoost} & \textbf{TabPFN} \\
        \midrule
        Accuracy  & $0.80 \pm 0.11$  & $0.84 \pm 0.10$ & $0.84 \pm 0.10$   & $0.82 \pm 0.11$ \\
        Precision & $0.83 \pm 0.10$  & $0.85 \pm 0.10$ & $0.86 \pm 0.10$   & $0.83 \pm 0.11$ \\
        Recall    & $0.80 \pm 0.11$  & $0.84 \pm 0.10$ & $0.84 \pm 0.10$   & $0.82 \pm 0.11$ \\
        $F_1$     & $0.79 \pm 0.12$  & $0.84 \pm 0.10$ & $0.84 \pm 0.10$   & $0.81 \pm 0.11$ \\
        \bottomrule
    \end{tabular}
    \label{tab:val:metrics:binary_101}
\end{table}

\section{Feature Importance Analysis}
To understand the relative importance of different synthesis parameters in predicting EMT zeolite formation, we employed two complementary feature importance analysis methods: the mean decrease in impurity (MDI) method and SHAP (Shapley Additive Explanations) analysis.

MDI is a method used in tree-based models, like XGBoost, that quantifies how much a feature contributes to reducing the weighted impurity (e.g., entropy) in the decision tree. A higher importance value indicates that the feature is more effective at splitting the data into purer nodes, thus playing a more significant role in the model's predictions. While this method provides a global ranking of features, it can be biased towards high-cardinality continuous features.

SHAP, on the other hand, is a more robust, model-agnostic method based on cooperative game theory. It calculates the marginal contribution of each feature to the prediction for each individual data point, providing a more nuanced and reliable measure of feature importance. These two methods offer complementary perspectives, allowing for a more comprehensive understanding of the model's decision-making process.

\begin{figure}[h!]
    \centering
    \includegraphics[width=0.7\textwidth]{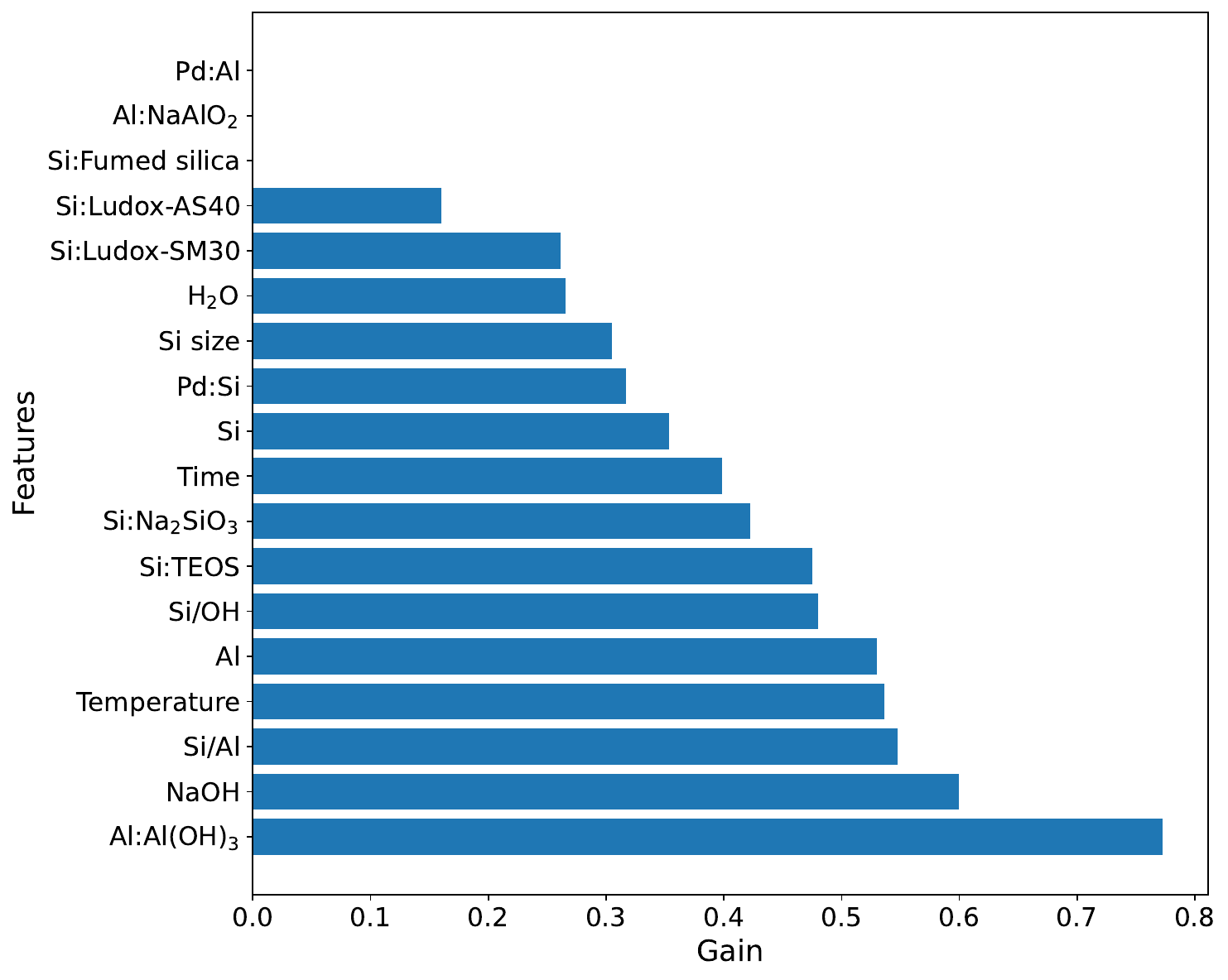}
    \caption{Synthesis parameter importance for EMT formation based on XGBoost MDI for the ternary classification model.}
    \label{fig:gains_ternary}
\end{figure}

The MDI analysis shows that the Si/OH ratio is a dominant feature, with an importance factor of approximately 0.5 (\fref{fig:gains_ternary}).
This finding is consistent with the SHAP analysis, which also identifies the Si/OH ratio as the most impactful feature. The source of aluminum, \ce{Al(OH)3}, also emerges as a significant factor, suggesting that the choice of precursor, not just the elemental ratios, influences EMT formation. Temperature and the Si/Al ratio show comparable importance values, highlighting their synergistic effect on crystallization dynamics.

\begin{figure}[h!]
    \centering
    \includegraphics[width=0.7\textwidth]{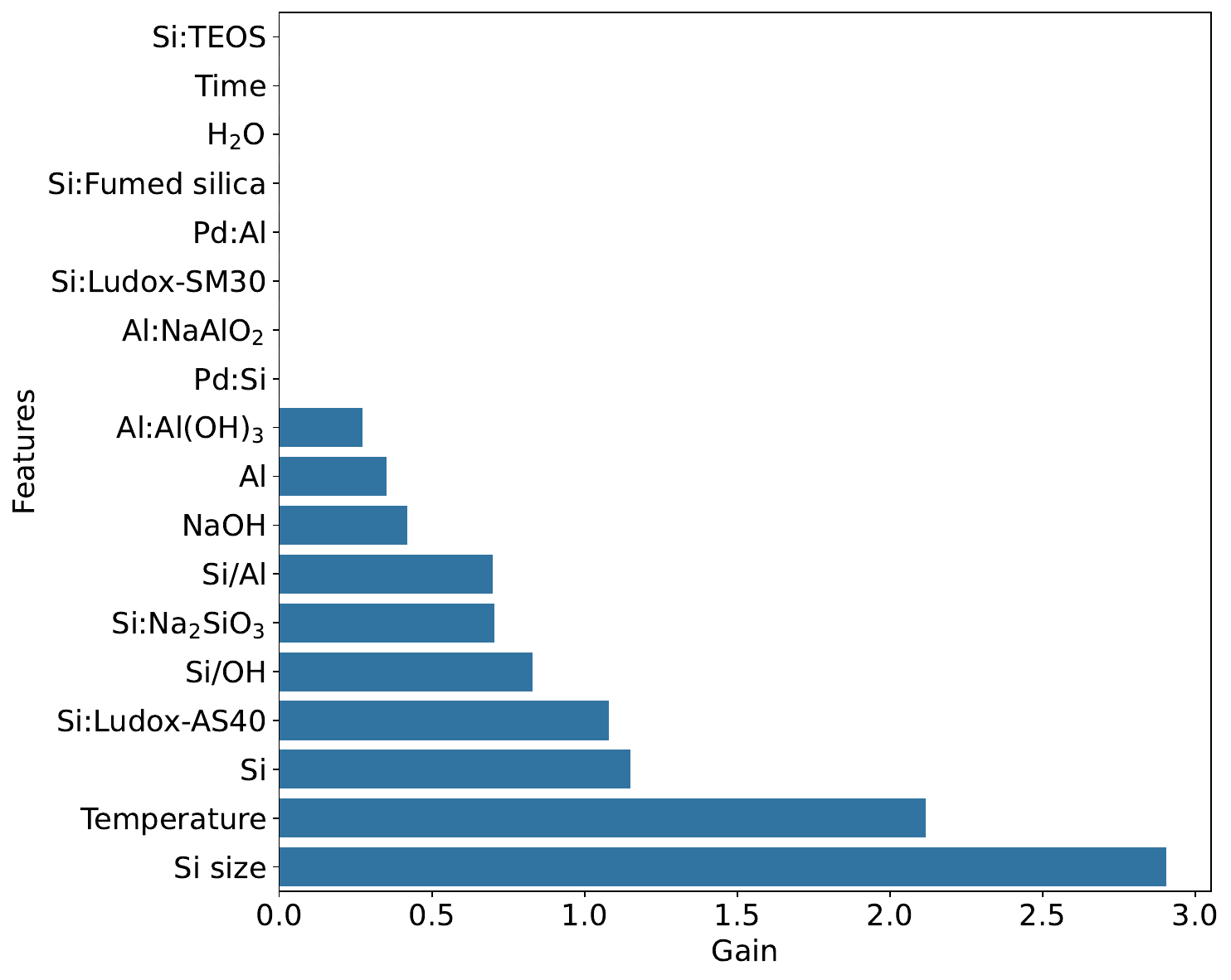}
    \caption{Synthesis parameter importance for EMT formation based on XGBoost MDI for the binary classification model.}
    \label{fig:gains_binary}
\end{figure}

\section{Evaluation time}
To illustrate the challenge of exhaustively exploring the synthesis parameter space, we can consider a simple combinatorial analysis. If we sample just 20 values for each of the 13 features, the total number of candidate synthesis conditions would be $20^{13} = 8.192 \times 10^{16}$.

The trained XGBoost model takes approximately $6 \times 10^{-5}$ seconds to make a single prediction, evaluating all these candidate conditions would require approximately $4.9 \times 10^{12}$ seconds, or about 155,000 years. This calculation highlights the impracticality of a brute-force approach and underscores the necessity of a more intelligent strategy. Machine learning models, by learning the underlying relationships between synthesis parameters and outcomes, can efficiently guide the exploration of this vast parameter space, enabling the identification of promising synthesis conditions in a fraction of the time.

\section{Post-processing of candidate synthesis conditions}
To down-select candidates from the initial set of 374 predicted synthesis conditions, we applied a series of post-processing steps prior to proposing conditions for experimental validation.
First, we resolved categorical features associated with the silicon sources.
For example, silicon sources are encoded as categorical variables with five possible options, where one source takes a value of 1, and the remaining sources take values of 0 (see S2 for details).
After inverse transformation of the model outputs, we retained only candidates for which the difference between the largest and second-largest silicon source values exceeded 0.6.
In these cases, we assigned the silicon source with the highest value to 1 and set all other sources to 0.
Candidates that did not meet this criterion (e.g., cases with values of 0.6 and 0.4 for two silicon sources) were discarded.
A similar rounding procedure was applied to all remaining binary categorical features with values close to 1 (e.g., 0.9) assigned to 1 and all other values set to 0.
Next, we adjusted continuous features representing discrete quantities. When the predicted value of the silicon precursor size was close to an integer, we rounded it to the nearest whole number.
For example, a predicted precursor size of 1.98 was replaced with 2.
Finally, after applying these manual adjustments, we recomputed the predicted probabilities and dissimilarity scores following the procedure described in the main text.
Finally, candidates with a minimum ``dissimilarity index" of 0.5 were retained.
From the resulting 28 candidates, we selected the six conditions with the highest predicted probabilities for experimental validation, which are discussed in the main text.
The remaining 22 candidates are reported in \tref{tab:refined_synthesis}.

\begin{table}[h!]
    \scriptsize
    \centering
    \caption{Remaining synthesis conditions proposed by the ML pipeline. The following features are the same for all samples: Si size (2), \ce{H2O} (450).}
    \label{tab:refined_synthesis}
    \begin{tabular}{ccccccccccccc}
        \toprule
        NaOH  & Al   & Si   & Pd:Si & Pd:Al & Si/OH & Si/Al & Time (hours) & Temperature ($^{\circ}$C) & Si:X         & Al:Y         & Probability & Dissimilarity \\
        \midrule
        36.50 & 1.84 & 6.03 & Yes   & Yes   & 0.17  & 3.58  & 72.71        & 41.24                     & \ce{Na2SiO3} & \ce{NaAlO2}  & 0.60        & 1.47          \\
        36.44 & 1.17 & 6.77 & Yes   & Yes   & 0.19  & 6.07  & 71.50        & 40.32                     & \ce{Na2SiO3} & \ce{NaAlO2}  & 0.61        & 2.02          \\
        36.47 & 0.61 & 8.35 & Yes   & Yes   & 0.23  & 8.92  & 69.86        & 38.72                     & \ce{Na2SiO3} & \ce{NaAlO2}  & 0.62        & 1.66          \\
        36.96 & 0.70 & 5.72 & Yes   & Yes   & 0.16  & 6.54  & 70.68        & 40.61                     & \ce{Na2SiO3} & \ce{NaAlO2}  & 0.62        & 1.98          \\
        36.50 & 0.94 & 7.98 & Yes   & Yes   & 0.22  & 7.67  & 70.46        & 39.18                     & \ce{Na2SiO3} & \ce{NaAlO2}  & 0.64        & 2.52          \\
        36.04 & 2.00 & 5.62 & Yes   & Yes   & 0.16  & 2.78  & 73.83        & 41.96                     & \ce{Na2SiO3} & \ce{NaAlO2}  & 0.66        & 7.86          \\
        36.79 & 1.54 & 7.44 & Yes   & Yes   & 0.20  & 5.58  & 71.11        & 39.73                     & \ce{Na2SiO3} & \ce{NaAlO2}  & 0.67        & 5.52          \\
        36.48 & 2.48 & 7.12 & Yes   & Yes   & 0.19  & 2.65  & 73.01        & 40.68                     & \ce{Na2SiO3} & \ce{NaAlO2}  & 0.69        & 1.34          \\
        36.73 & 0.78 & 5.52 & Yes   & Yes   & 0.15  & 6.14  & 71.24        & 40.97                     & \ce{Na2SiO3} & \ce{NaAlO2}  & 0.69        & 1.41          \\
        36.71 & 2.40 & 7.32 & Yes   & Yes   & 0.20  & 3.05  & 72.45        & 40.32                     & \ce{Na2SiO3} & \ce{NaAlO2}  & 0.71        & 2.00          \\
        36.45 & 2.15 & 7.49 & Yes   & Yes   & 0.20  & 3.90  & 72.41        & 40.22                     & \ce{Na2SiO3} & \ce{NaAlO2}  & 0.71        & 2.69          \\
        35.90 & 2.20 & 6.24 & Yes   & Yes   & 0.17  & 2.74  & 73.96        & 41.64                     & \ce{Na2SiO3} & \ce{NaAlO2}  & 0.72        & 2.13          \\
        36.79 & 2.51 & 8.16 & Yes   & Yes   & 0.22  & 3.41  & 72.02        & 39.63                     & \ce{Na2SiO3} & \ce{NaAlO2}  & 0.72        & 2.49          \\
        36.76 & 2.10 & 5.86 & Yes   & Yes   & 0.16  & 2.73  & 72.75        & 41.33                     & \ce{Na2SiO3} & \ce{NaAlO2}  & 0.73        & 7.17          \\
        36.04 & 1.03 & 4.90 & Yes   & Yes   & 0.14  & 4.95  & 72.93        & 42.06                     & \ce{Na2SiO3} & \ce{NaAlO2}  & 0.78        & 1.60          \\
        36.53 & 2.26 & 8.32 & Yes   & Yes   & 0.23  & 4.26  & 71.97        & 39.54                     & \ce{Na2SiO3} & \ce{NaAlO2}  & 0.79        & 2.67          \\
        36.56 & 2.59 & 7.95 & Yes   & Yes   & 0.22  & 3.02  & 72.58        & 39.99                     & \ce{Na2SiO3} & \ce{NaAlO2}  & 0.81        & 2.89          \\
        43.10 & 2.14 & 5.74 & No    & No    & 0.14  & 2.60  & 63.44        & 37.25                     & Ludox-AS40   & \ce{Al(OH)3} & 0.81        & 6.05          \\
        36.90 & 1.90 & 5.23 & Yes   & Yes   & 0.14  & 2.77  & 72.62        & 41.66                     & \ce{Na2SiO3} & \ce{NaAlO2}  & 0.84        & 0.73          \\
        36.87 & 1.57 & 5.61 & Yes   & Yes   & 0.15  & 4.01  & 72.02        & 41.20                     & \ce{Na2SiO3} & \ce{NaAlO2}  & 0.84        & 1.95          \\
        36.67 & 1.01 & 4.31 & Yes   & Yes   & 0.12  & 4.54  & 72.28        & 42.12                     & \ce{Na2SiO3} & \ce{NaAlO2}  & 0.85        & 0.59          \\
        36.82 & 2.85 & 7.78 & Yes   & Yes   & 0.21  & 2.17  & 72.62        & 40.09                     & \ce{Na2SiO3} & \ce{NaAlO2}  & 0.86        & 1.98          \\
        \bottomrule
    \end{tabular}
\end{table}

\section{Experimental Synthesis Protocol}

The experimental range was expanded by varying the precursors, temperature, and synthesis time. The growth mixture for zeolite EMT was prepared with molar compositions of \( x\,\mathrm{SiO_2} : y\,\mathrm{Al_2O_3} : z\,\mathrm{NaOH} : 450\,\mathrm{H_2O} \).

The synthesis was carried out as follows:
\begin{enumerate}
    \item Solution A was prepared by dissolving sodium aluminate and sodium hydroxide in deionized (DI) water in a polypropylene bottle, with stirring in an ice bath until the mixture was homogeneous.
    \item Solution B was prepared by adding one of the silica sources (sodium silicate, fumed silica, or tetraethyl orthosilicate (TEOS)) to DI water in a separate polypropylene bottle. Sodium silicate was used as the nominal silica source unless otherwise stated.
    \item Solution B was stirred in an ice bath until homogeneous, then added to Solution A. The resulting growth mixture was stirred in the ice bath for an additional 10 minutes.
    \item The mixture was heated under static conditions in a Thermo Fisher Precision oven at a temperature between 35 and 45~\(^\circ\)C.
    \item The bottles were removed from the oven at various time intervals between 3 and 72~hr and cooled to 25~\(^\circ\)C in a water bath.
    \item The solid product was recovered by three cycles of centrifugation (6000~rpm for 5~minutes using a Corning LSE centrifuge) and washing with DI water.
    \item The final product was dried in air at room temperature prior to characterization.
\end{enumerate}

\section{Characterization of ML-Guided EMT Syntheses}
\label{sec:characterization:ml:emt}

\paragraph{X-ray diffraction.}
Powder X-ray diffraction (PXRD) patterns of dried solids were collected on a Rigaku SmartLab diffractometer with a Cu K$\alpha$ source (40 kV, 30 mA).

\begin{figure}[h!]
    \centering
    \includegraphics[width=0.5\textwidth]{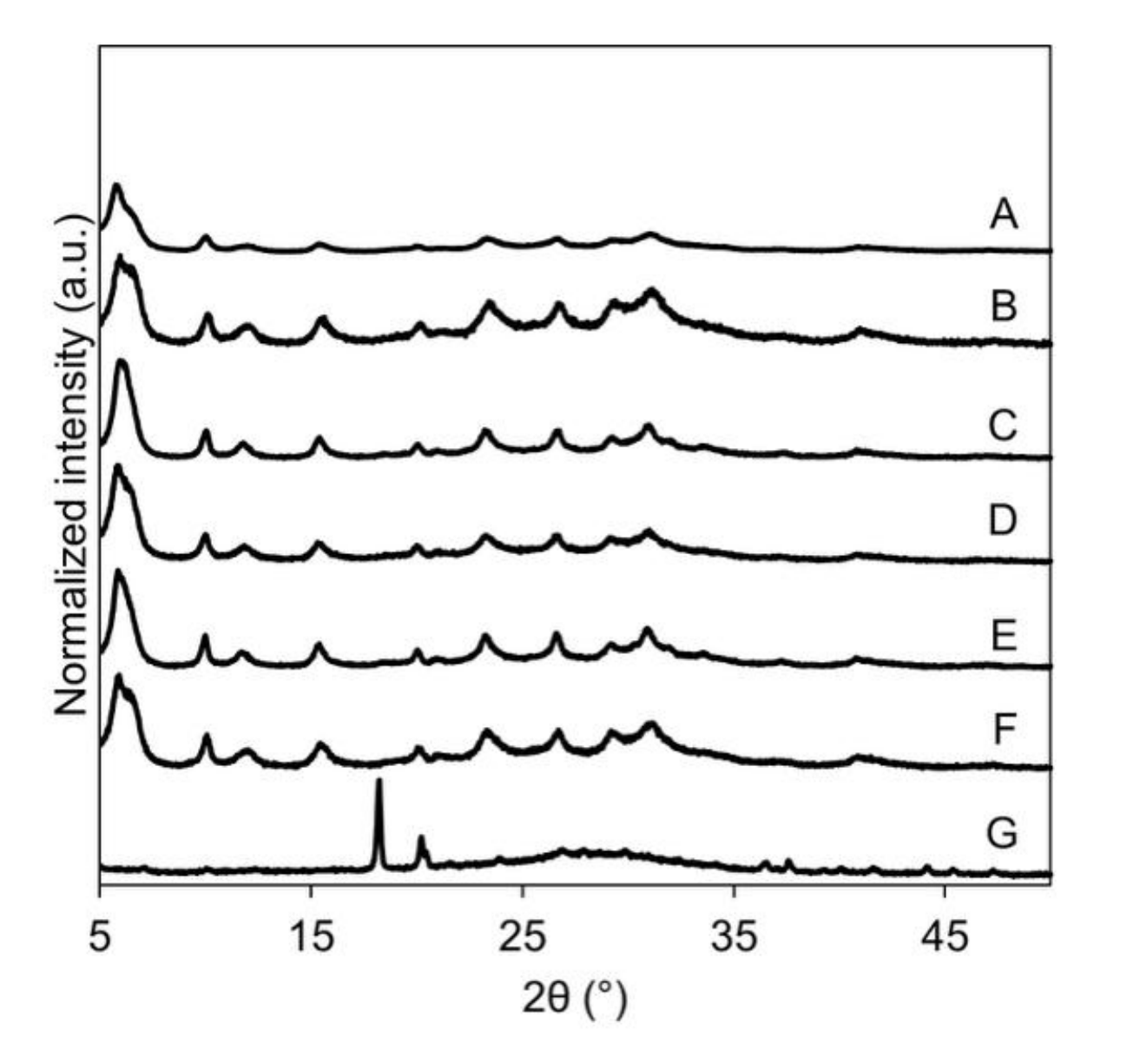}
    \caption{XRD patterns of (A) reference EMT from the IZA database and (B)--(G) EMT synthesized under the conditions listed in Table~IV of the main text.
        (B) \ce{36.59 NaOH} : \ce{2.93 Al(OH)3} : \ce{7.58 SiO2} : \ce{450 H2O}.
        (C) \ce{36.93 NaOH} : \ce{1.26 Al(OH)3} : \ce{4.15 SiO2} : \ce{450 H2O}.
        (D) \ce{36.65 NaOH} : \ce{1.65 Al(OH)3} : \ce{5.40 SiO2} : \ce{450 H2O}.
        (E) \ce{36.76 NaOH} : \ce{1.12 Al(OH)3} : \ce{5.14 SiO2} : \ce{450 H2O}.
        (F) \ce{37.39 NaOH} : \ce{2.08 Al(OH)3} : \ce{5.27 SiO2} : \ce{450 H2O}.
        (G) \ce{36.99 NaOH} : \ce{1.04 Al(OH)3} : \ce{5.35 SiO2} : \ce{450 H2O}.}
    \label{fig:xrd_patterns}
\end{figure}

\paragraph{Scanning electron microscopy.}
Scanning electron microscopy (SEM) images were obtained using an FEI-235 Dual-Beam Focused Ion Beam instrument operated at 15 kV and a 5 mm working distance.

\begin{figure}[h!]
    \centering
    \includegraphics[width=0.5\textwidth]{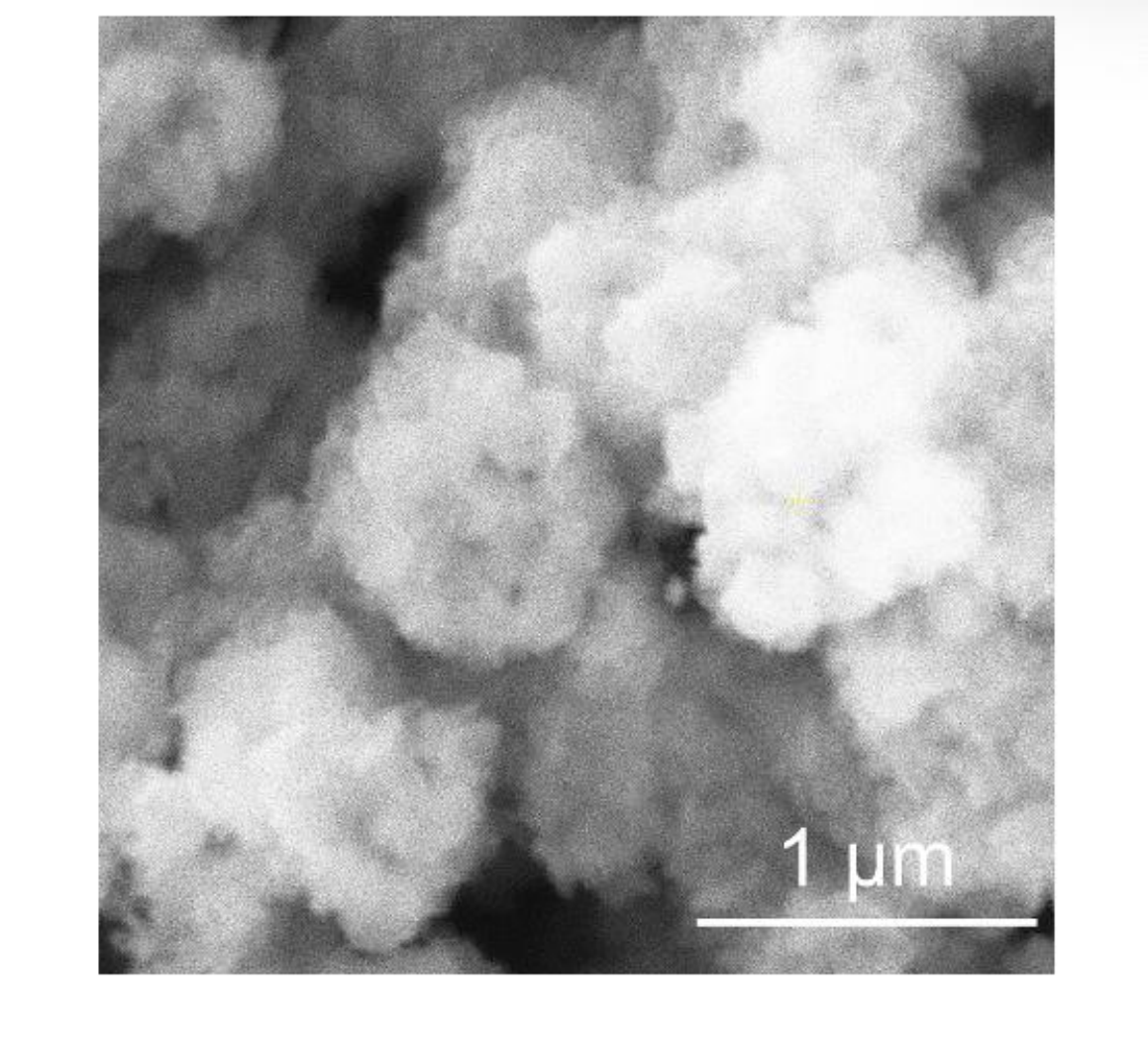}
    \caption{SEM image of EMT synthesized with a molar ratio of \ce{36.9 NaOH} : \ce{1 Al(OH)3} : \ce{4 SiO2} : \ce{450 H2O}, using \ce{NaAlO2} and \ce{Na2SiO3} as aluminum and silica sources. The mixture was heated at 40~\textdegree C for 72~h.}
    \label{fig:sem_images}
\end{figure}

\section{External Validation on Literature-Reported Syntheses}
\label{sec:experimental_dataset}

This section provides the methodological detail supporting the literature-validation analysis in Section~IIID of the main manuscript: how the literature dataset was compiled and how descriptors not reported in the original papers were imputed (i.e., filled in with substitute values so the model can still be evaluated).

\paragraph{Literature compilation and search strategy.}
The dataset was compiled from the EMT zeolite synthesis literature, yielding 16 pure-EMT hydrothermal synthesis entries.
Each entry corresponds to a unique gel composition reported in the source; when the same synthesis appeared in more than one publication, only one instance was kept to avoid duplication, and entries with uncertain phase assignments were excluded.
The dataset, together with the model's prediction for each entry, is listed in Table~\ref{tab:emt_dataset_pure_only}.

\paragraph{Handling of unreported descriptors.}
Several descriptors required by the model were not reported consistently in the original papers, including both numerical and categorical features.
Among the numerical descriptors, the \ce{SiO2} precursor particle size was not reported in any of the literature sources, and the crystallization time was missing in some entries (see Table~I of the main manuscript for the definitions).
The categorical descriptors most commonly missing were whether the Si and Al precursors were pre-dissolved before synthesis.
For the two categorical descriptors, we did not impute a single value; instead, when they were missing, we evaluated all possible combinations of the categorical states and averaged the resulting model predictions.
For numerical descriptors, we used mean imputation: missing values were replaced by the mean of that feature computed from the in-house training dataset.
For the \ce{SiO2} particle size, which was not reported in any of the literature sources, the in-house mean is 3.25 nm; to reduce dependence on a single imputed value, we duplicated each literature row using five \ce{SiO2} particle sizes (2, 3, 3.25, 4, and 5 nm) and report the prediction and dissimilarity averaged over the five evaluations.

Because the imputed values are drawn from the in-house training dataset, they may place literature cases closer to the in-house synthesis space than they would be if the true, unreported values were available.
The literature validation should therefore be viewed as an external evaluation using the reported literature descriptors together with explicitly documented assumptions for missing values, rather than a fully independent test across all model inputs.

\begin{table*}[htb!]
    \scriptsize
    \centering
    \caption{Experimental EMT synthesis conditions compiled from the literature and used for external validation of the model.
        Only literature cases reported as pure EMT hydrothermal syntheses are included.
        The \ce{SiO2} precursor particle size is not shown because it was not reported in any source and was imputed identically for all entries (see the imputation procedure described above).
        The ``Prediction'' column reports the model's classification (``Yes'' for predicted EMT, ``No'' for predicted non-EMT).
    }
    \label{tab:emt_dataset_pure_only}
    \resizebox{\textwidth}{!}{%
        \begin{tabular}{ccccccccccccc}
            \toprule
            NaOH  & Al   & Si    & \ce{H2O} & Si/OH & Si/Al & Time (h) & Temperature ($^{\circ}$C) & Si:X         & Al:Y        & Prediction & Dissimilarity & Source                      \\
            \midrule
            5.40  & 2.00 & 10.96 & 153.00   & 2.03  & 5.48  & 624.00   & 90.00                     & \ce{Na2SiO3} & \ce{NaAlO2} & Yes        & 38.61         & \cite{wendelbo1999tumbling} \\
            4.80  & 2.00 & 10.00 & 140.00   & 2.08  & 5.00  & 168.00   & 115.00                    & Ludox-AS40   & \ce{NaAlO2} & Yes        & 38.83         & \cite{wu1995synthesis}      \\
            4.00  & 2.00 & 10.00 & 140.00   & 2.50  & 5.00  & 240.00   & 115.00                    & Ludox-AS40   & \ce{NaAlO2} & Yes        & 44.77         & \cite{wu1995synthesis}      \\
            4.20  & 2.00 & 10.00 & 140.00   & 2.38  & 5.00  & 24.00    & 71.65                     & \ce{Na2SiO3} & \ce{NaAlO2} & No         & 39.54         & \cite{dougnier1995emt}      \\
            36.90 & 2.00 & 5.15  & 240.00   & 0.14  & 2.58  & 36.00    & 40.00                     & \ce{Na2SiO3} & \ce{NaAlO2} & Yes        & 1.48          & \cite{georgieva2015control} \\
            34.00 & 2.00 & 5.15  & 250.00   & 0.15  & 2.58  & 24.00    & 40.00                     & \ce{Na2SiO3} & \ce{NaAlO2} & No         & 2.10          & \cite{georgieva2015control} \\
            34.00 & 2.00 & 5.15  & 350.00   & 0.15  & 2.58  & 36.00    & 40.00                     & \ce{Na2SiO3} & \ce{NaAlO2} & Yes        & 3.84          & \cite{georgieva2015control} \\
            34.00 & 2.00 & 5.15  & 350.00   & 0.15  & 2.58  & 8.00     & 60.00                     & \ce{Na2SiO3} & \ce{NaAlO2} & Yes        & 5.53          & \cite{georgieva2015control} \\
            34.00 & 2.00 & 5.15  & 550.00   & 0.15  & 2.58  & 84.00    & 40.00                     & \ce{Na2SiO3} & \ce{NaAlO2} & Yes        & 3.72          & \cite{georgieva2015control} \\
            34.00 & 2.00 & 5.15  & 450.00   & 0.15  & 2.58  & 12.00    & 60.00                     & \ce{Na2SiO3} & \ce{NaAlO2} & Yes        & 4.19          & \cite{georgieva2015control} \\
            34.00 & 2.00 & 5.15  & 450.00   & 0.15  & 2.58  & 20.00    & 40.00                     & \ce{Na2SiO3} & \ce{NaAlO2} & Yes        & 1.78          & \cite{georgieva2015control} \\
            34.00 & 2.00 & 5.15  & 450.00   & 0.15  & 2.58  & 48.00    & 40.00                     & \ce{Na2SiO3} & \ce{NaAlO2} & Yes        & 1.22          & \cite{georgieva2015control} \\
            34.00 & 2.00 & 5.15  & 450.00   & 0.15  & 2.58  & 36.00    & 40.00                     & \ce{Na2SiO3} & \ce{NaAlO2} & Yes        & 1.44          & \cite{georgieva2015control} \\
            34.00 & 2.00 & 5.15  & 450.00   & 0.15  & 2.58  & 24.00    & 40.00                     & \ce{Na2SiO3} & \ce{NaAlO2} & No         & 1.69          & \cite{georgieva2015control} \\
            36.90 & 2.00 & 5.15  & 240.30   & 0.14  & 2.58  & 36.00    & 30.00                     & \ce{Na2SiO3} & \ce{NaAlO2} & No         & 2.88          & \cite{ng2012capturing}      \\
            36.00 & 2.00 & 5.00  & 217.00   & 0.14  & 2.50  & 28.00    & 28.00                     & \ce{Na2SiO3} & \ce{NaAlO2} & No         & 3.59          & \cite{ng2015emt}            \\
            \bottomrule
        \end{tabular}%
    }
    \label{tab:emt_dataset}
\end{table*}

\clearpage
%